\documentclass[aps,prl,twocolumn,floatfix,showpacs,superscriptaddress]{revtex4-1}
\usepackage{graphicx,amsmath,amsfonts,mathrsfs}
\usepackage{longtable}
\usepackage{wasysym}
\usepackage{trsym}
\usepackage{marvosym}

\begin{document}

\title{Localised Magnetism in 2D Electrides }

\author{Danis I  Badrtdinov}
\affiliation{Department of Theoretical Physics and Applied Mathematics, \\ Ural Federal University, Mira St.~19, 620002 Yekaterinburg, Russia}
\author{Sergey A  Nikolaev}
\email{nikolaev.s.aa@m.titech.ac.jp}
\affiliation{Institute of Innovative Research, Tokyo Institute of Technology, \\ 4259 Nagatsuta, Midori, Yokohama 226-8503, Japan}
\affiliation{Department of Theoretical Physics and Applied Mathematics, \\ Ural Federal University, Mira St.~19, 620002 Yekaterinburg, Russia}

\begin{abstract}
In this work, we investigate intrinsic magnetic properties of monolayer electrides LaBr$_{2}$ and La$_{2}$Br$_{5}$, where excess electrons do not reside at any atomic orbital and act as anions located at interstitial regions. Having demonstrated that conventional first-principles approaches are incapable of treating such non-atomic magnetic orbitals largely underestimating insulating band gaps, we construct effective electronic models in the basis of Wannier functions associated with the anionic states to unveil the microscopic mechanism underlying magnetism in these systems. Being confined at zero-dimensional cavities in the crystal structure, the anionic electrons will be shown to reveal an exotic duality of strong localisation like in $d$- and $f$-electron systems and large spatial extension inherent to delocalised atomic orbitals. While the former tends to stabilise a Mott-insulating state with localised magnetic moments, the latter results in direct exchange between neighbouring anionic electrons that dominates over antiferromagnetic superexchange interactions. On the basis of the derived spin models, we argue that any long-range magnetic order is prohibited in LaBr$_{2}$ by Mermin-Wagner theorem, while intersite anisotropy in La$_{2}$Br$_{5}$ stabilises weakly coupled ferromagnetic chains along the monoclinic $\mathbf{b}$ axis.  Our study shows that electride materials combining peculiar features of both localised and delocalised atomic states constitute a unique class of strongly correlated materials.
\end{abstract}

\maketitle

\par \emph{Introduction}. Electrides are a unique class of ionic materials, where the electron density is neither localised at any atomic orbital, nor fully delocalised like in metals. Instead, their electrons occupy interstitial regions formed by cavities in the crystal structure, where they act as anions~\cite{dye1,dye2}. Materials with anionic electrons offer versatile functionalities, such as high electrical conductivity~\cite{matsuishi2}, ultra-low work functions~\cite{toda}, and non-linear optical responses~\cite{sguan}, ranging their applications as electron emitters~\cite{swkim}, battery anodes~\cite{junping}, and agent catalysts~\cite{kitano}.

\par Their physical properties are in large part determined by topology of the voids confining anionic electrons. In 2003, Matsuishi \emph{et al}~\cite{matsuishi1} demonstrated the first inorganic material [Ca$_{24}$Al$_{28}$O$_{64}$]$^{+4}$$\cdot$4e$^{-}$ stable at room temperature, which shows the high density of anionic electrons trapped in the crystallographic cages, forming a quasi-zero-dimensional electride with high electronic conductivity due to tunnelling through the cages~\cite{hosono1,hosono2}. Recently, a novel layered electride [Ca$_{2}$N]$^{+}\cdot $e$^{-}$ has been reported, where owing to a highly delocalised nature the quasi-two-dimensional anionic electrons confined in the interlayer regions display an extreme electron mobility and long mean scattering time~\cite{kimoonlee,wonnoh}. Later on, strong localisation of anionic electrons has been observed in a layered transition-metal hypocarbide [Y$_{2}$C]$^{2+}$$\cdot$2e$^{-}$ with the quasi-two-dimensional anionic electrons in the interlayer spaces~\cite{y2celectride,y2c2}. Finally, it was shown that Ca$_{2}$N can be exfoliated into two-dimensional (2D) nanosheets by liquid exfoliation, while preserving anionic electrons on its surface~\cite{electrene1,electrene2}. Nowadays, 2D electrides, or electrenes, is a special topic of research fusing the concepts of electride chemistry and 2D materials \cite{electrenereview}. 

\par In this regards, 2D materials are of extreme interest featuring various exceptional properties and applications~\cite{graphene,mos2}. Of particular importance is intrinsic magnetism in atomically thin systems, where over the years various prototype examples have been reported to exhibit magnetic properties~\cite{chengong,bevibhuang,yujuin,bonilla,xinghan}. While all these materials contain $d$-electron elements that form localised magnetic moments in a half-filled electronic shell, a natural question is whether such magnetism can be realised in 2D electrides. 

\par Indeed, when trapped in zero-dimensional cavities, the half-filled anionic states can be subject to on-site Coulomb repulsion as a result of their geometrical confinement. For example, a Mott-insulating phase has been reported in $\alpha$- and $\beta$-Yb$_{5}$Sb$_{3}$~\cite{mottelectride}, where the anionic electrons residing in zero- and one-dimensional interstitial regions manifest semiconducting conductivity and a Curie-type magnetism due to localisation of the electron moment. Nevertheless, realisation of such properties in 2D or layered electrides has not yet been confirmed. 

\par High-throughput calculations have become a powerful tool for searching novel electrides, extending the number of possible candidates well beyond the formerly known materials \cite{burton,search1,search2,search3}. Upon screening a database of the Materials Project~\cite{material}, which includes first-principles data for a large set of known compounds, several layered electrides have been identified that feature both zero- and one-dimensional anionic electrons and may be promising candidates for advanced two-dimensional materials~\cite{zhou}. Among them, metal halides in the MoS$_{2}$-like structure~\cite{labr2} were found to be magnetic with the anionic electrons localised at the center of metal-halogen hexagons. However, the microscopic origin of magnetism in these systems has not been explained.

\begin{figure}
\begin{center}
\includegraphics[width=0.50\textwidth]{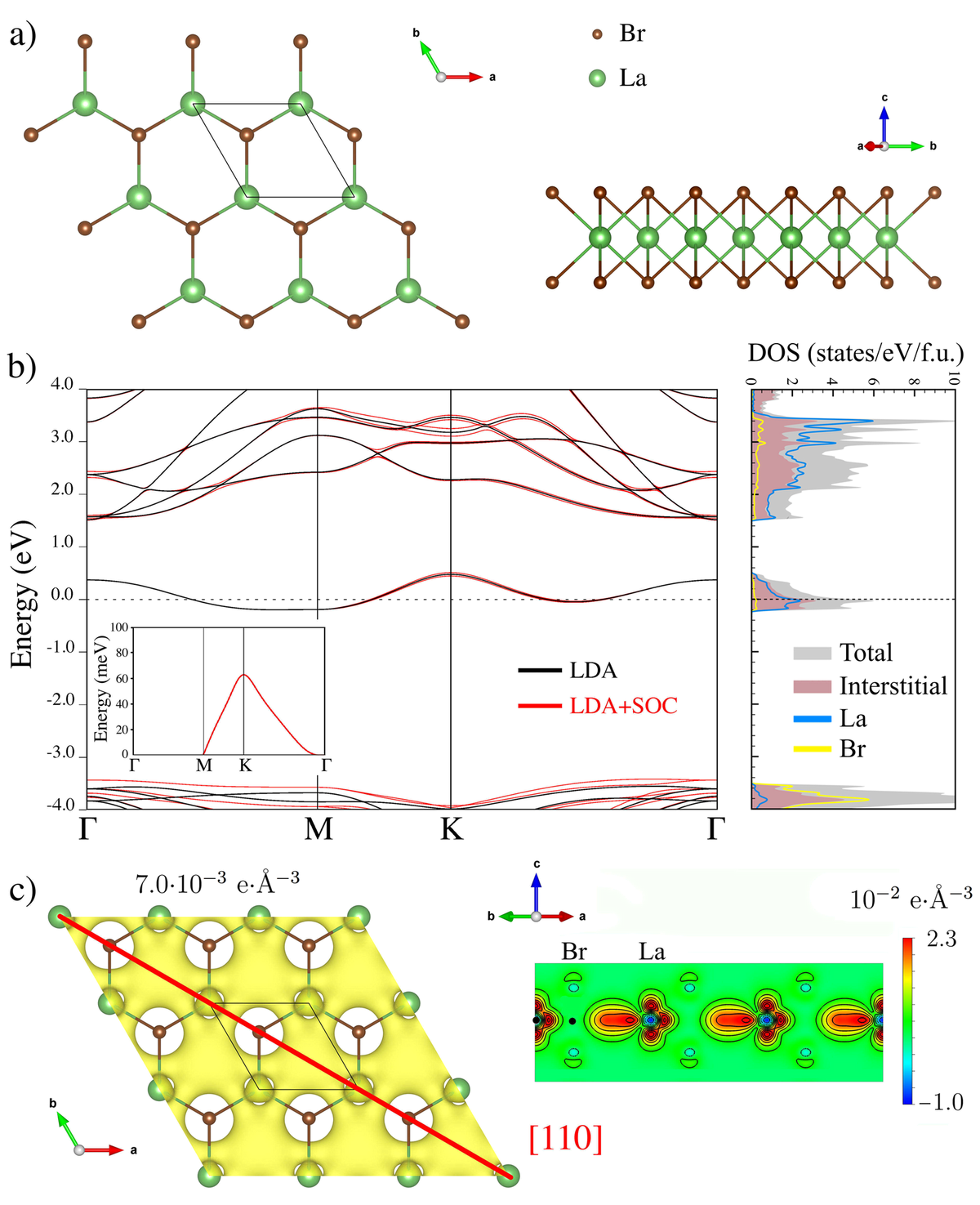}
\end{center}
\caption{a) Crystal structure of a single monolayer LaBr$_{2}$ (visualised by \texttt{VESTA}~\cite{vesta}); b) Band structures calculated by using LDA with and without spin-orbit coupling (LDA+SOC and LDA, respectively). The inset shows the SOC energy splitting of the band at the Fermi level; c) Charge density corresponding to the band at the Fermi level and its projection onto the [110] plane.}
\label{fig:band1}
\end{figure}

\begin{figure}
\begin{center}
\includegraphics[width=0.50\textwidth]{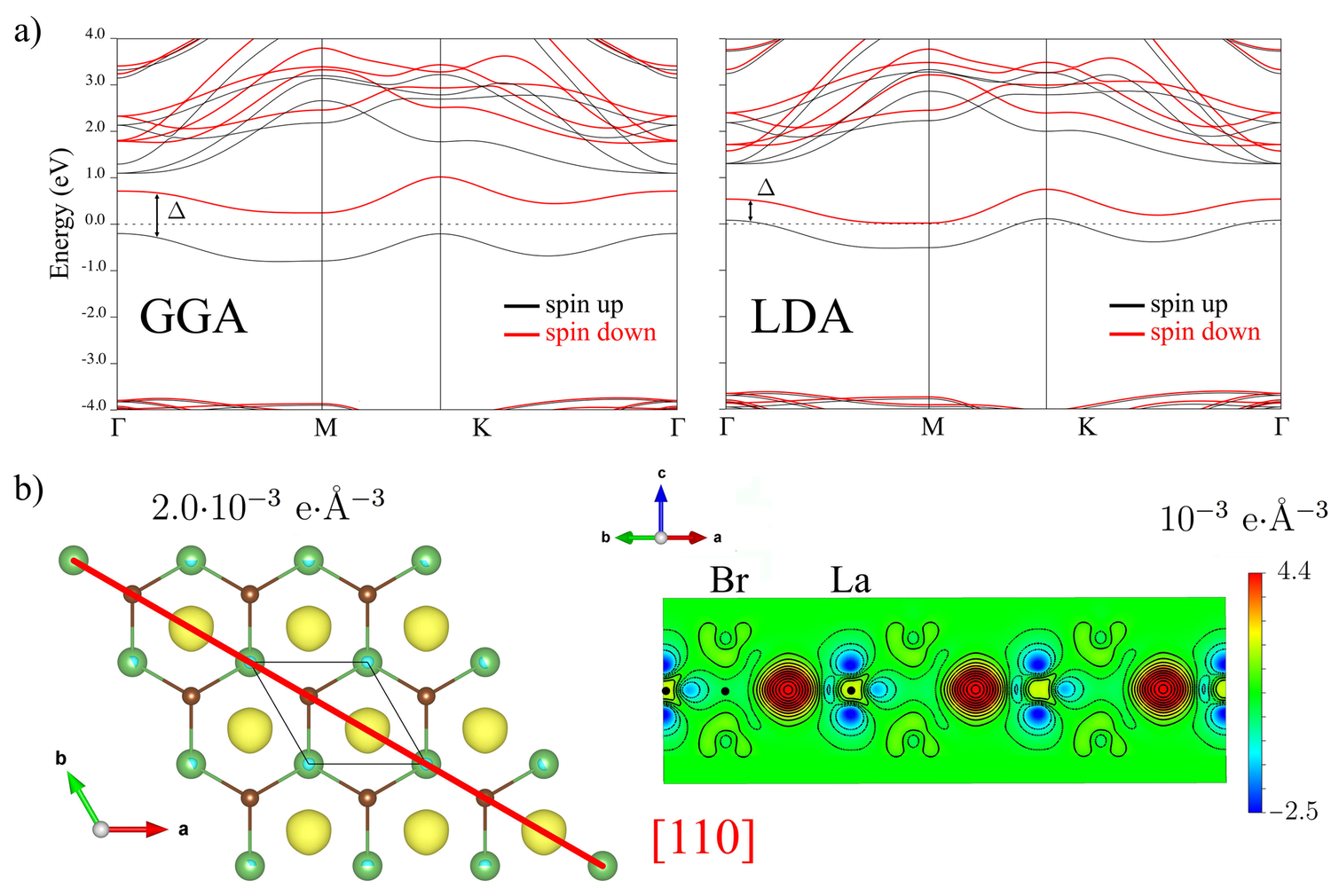}
\end{center}
\caption{a) Band structures of the ferromagnetic state in LaBr$_{2}$ calculated within GGA and LDA; b) Magnetisation density of the anionic electron as obtained from GGA for the ferromagnetic state of LaBr$_{2}$. }
\label{fig:sp}
\end{figure}

\par Motivated by these studies, we combine rigorous first-principles calculations and low-energy electronic models to investigate intrinsic magnetism in electrenes LaBr$_{2}$ and La$_{2}$Br$_{5}$ and argue that the anionic electrons in these systems feature a dual nature of strong localisation and large spatial extension. As a result, the former will allow to exhibit a Mott-insulating phase with localised magnetic moments at non-atomic orbitals, whereas the latter will promote dominant direct exchange between neighbouring anionic states stabilising a ferromagnetic order. Our study shows that a wide class of strongly correlated materials can be enriched with electrides, whose half-filled anionic states are confined at zero-dimensional cavities.
 
\par \emph{First-principles calculations}. Electronic structure calculations were performed within local density approximation (LDA) \cite{kohnsham,lda} and generalised gradient approximation (GGA) \cite{ggapbe} for the exchange-correlation functional by using the projected augmented wave formalism \cite{paw}, as implemented in the Vienna ab-initio simulation package \texttt{VASP}~\cite{vasp}, norm-conserving pseudopotentials, as implemented in the \texttt{Quantum ESPRESSO} package~\cite{qe}, and full-potential linearised augmented plane wave method, as implemented in the \texttt{Elk} code~\cite{elk}. The calculation details are given in Supplementary Material~\cite{supp}.

\begin{figure}
\begin{center}
\includegraphics[width=0.50\textwidth]{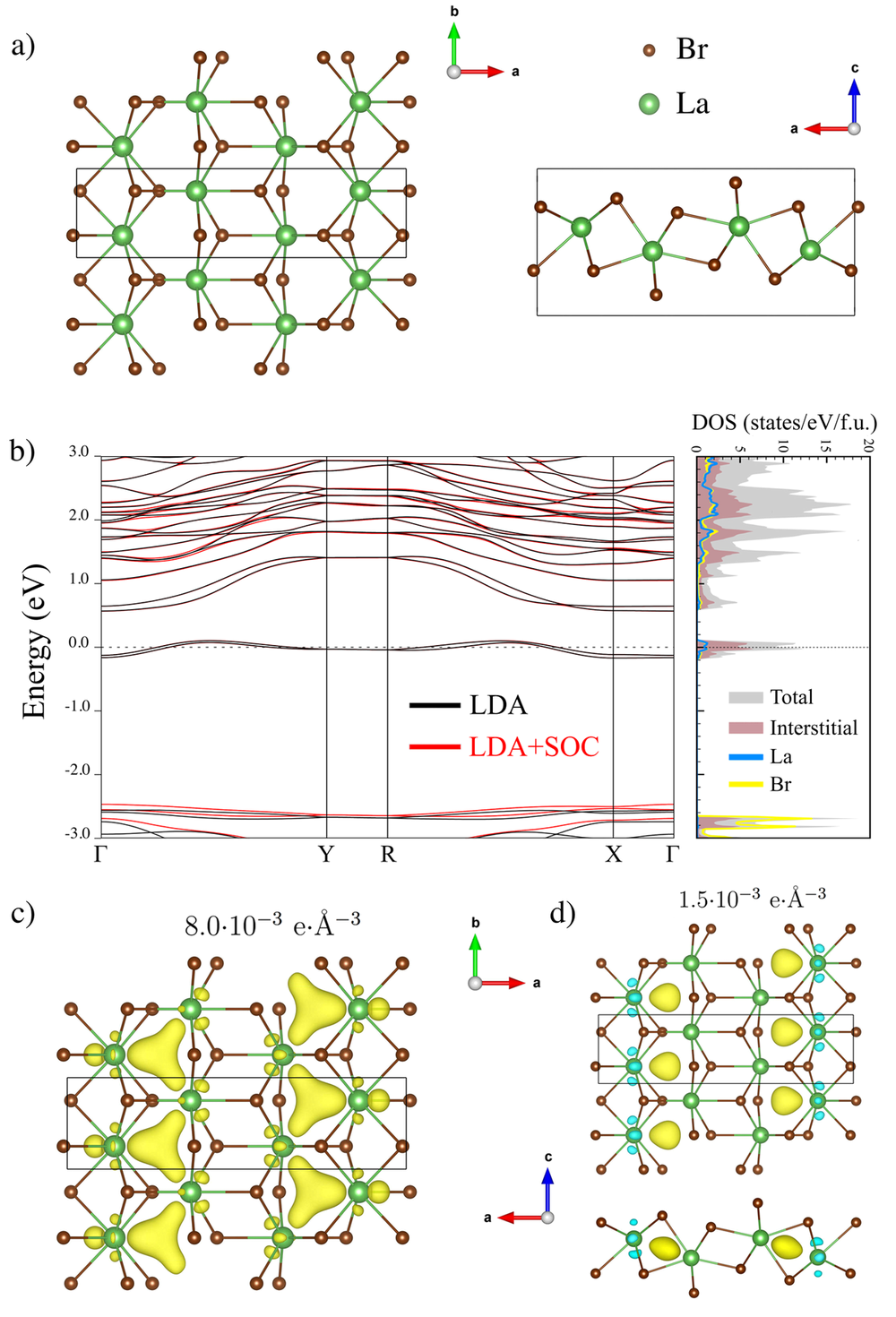}
\end{center}
\caption{a) Crystal structure of a single monolayer La$_{2}$Br$_{5}$ (visualised by \texttt{VESTA}~\cite{vesta}); b) Band structures calculated by using LDA with and without spin-orbit coupling (LDA+SOC and LDA, respectively). c) Charge and d) magnetisation densities of the anionic electrons as obtained from GGA for the ferromagnetic state of La$_{2}$Br$_{5}$.}
\label{fig:band2}
\end{figure}

\par Experimentally synthesised by lithium reduction of lanthanum tribromide, LaBr$_{2}$ has a layered $P6_{3}/mmc$ structure with the La atoms sandwiched between the Br layers and resembles the 2H phase of MoS$_{2}$~\cite{labr2}. It was shown in previous studies~\cite{zhou} that the calculated exfoliation energy allows to isolate single monolayers, similar to other layered electride materials, such as Ca$_{2}$N. Results of the electronic structure calculations for monolayer LaBr$_{2}$ shown in figure~\ref{fig:band1} reveal a single band at the Fermi level which is mainly formed by the La $d$ states and anionic electrons from interstitial regions. Indeed, the charge density corresponding to this state indicates that its significant portion is confined at the hexagon cavity and can be attributed to the excess electron as [La$^{+3}$Br$^{-}_{2}$]$\cdot$e$^{-}$. The band at the Fermi level is further split when spin-orbit coupling is taking into account, but owing to the loosely bound nature of anionic electrons it remains weakly dispersive.

\par Located at the Fermi level, this state features strong Stoner instability allowing for a formation of spontaneous magnetisation~\cite{stoner}, and the spin-polarised calculations show a ferromagnetic state with the magnetisation density strictly localised at the hexagon cavity, as shown in figure~\ref{fig:sp}. This fact clearly demonstrates the localised nature of anionic electrons in LaBr$_{2}$. Importantly, the corresponding spitting of the spin-up and spin-down states $\Delta$ strongly varies within LDA and GGA from 0.46 eV to 0.92 eV, respectively, thus indicating a crucial role of electronic correlation effects. 

\par A similar situation is observed in La$_{2}$Br$_{5}$ with a monoclinic $P2_{1}m$ structure, whose single crystals are also obtained by lithium reduction of lanthanum tribromide in sealed tantalum ampoules~\cite{labr2}. As seen from figure~\ref{fig:band2}, monolayer La$_{2}$Br$_{5}$ has two half-filled anionic states located at the Fermi level, that due to Stoner instability form magnetic moments residing at the center of each hexagon in the unit cell. Hence, our preliminary calculations certainly evidence the presence of magnetic moments in LaBr$_{2}$ and La$_{2}$Br$_{5}$ formed by non-atomic excess electrons and localised at interstitial areas.

\par \emph{Electronic models}. When confined at a zero-dimensional cavity, the half-filled anionic states can experience strong on-site Coulomb repulsion maximising their correlation effects and leading to a Mott-insulating state with localised magnetic moments, similar to what happens in many transition-metals oxides~\cite{imada}. However, treating such systems in the framework of conventional ab-initio calculations may be questionable. While DFT+$U$ calculations come in handy to describe the features of strongly correlated $d$- and $f$-electron systems~\cite{dftu}, they are not suitable for electrides since the on-site $U$ cannot be directly introduced on any states but atomic orbitals. On the other hand, using semi-empirical hybrid functionals~\cite{hse} does not allow to unveil the underlying mechanism stabilising magnetism. To capture such effects in electrides, one has to go beyond these approaches and make use of low-energy electronic models constructed solely for the anionic states. In this work, we consider the following one-orbital extended Hubbard model for each system in question:
\begin{equation}
\begin{aligned}
\mathcal{H}&=\sum\limits_{\langle ij \rangle,\sigma\sigma'}t_{ij}^{\sigma\sigma'}c_{i}^{\dagger\sigma}c_{j}^{\!\!\phantom{\dagger}\sigma'}+\sum_{i}U_{i}^{}n_{i}^{\uparrow}n_{i}^{\downarrow} \\
&+\sum\limits_{\langle ij \rangle,\sigma}V^{}_{ij}n^{\sigma}_{i}n_{j}^{\bar{\sigma}} + \sum\limits_{\langle ij \rangle,\sigma}(V^{}_{ij}-J^{F}_{ij})n^{\sigma}_{i}n_{j}^{\sigma},
\end{aligned}
\end{equation}
\noindent where $c^{\dagger\sigma}_{i}$ ($c^{\!\!\phantom{\dagger}\sigma}_{i}$) creates (annihilates) an electron with spin $\sigma$ at site $i$, $n_{i}=c^{\dagger\sigma}_{i}c^{\!\!\phantom{\dagger}\sigma}_{i}$ is the density operator, $t_{ij}^{\sigma\sigma'}$ is the hopping parameter including spin-orbit coupling (given as complex numbers), and the interaction terms include the on-site Coulomb $U$, intersite Coulomb $V$, and intersite direct exchange $J^{F}$ interactions. This model is formulated in the basis of Wannier functions for the magnetically active states near the Fermi level by using the procedure of maximal localisation as implemented in the \texttt{wannier90} package~\cite{wanapp,wan90}. The corresponding hopping parameters are obtained by projecting the LDA Hamiltonian with spin-orbit coupling onto the basis of Wannier functions, and the matrix elements of Coulomb and direct exchange interactions were calculated within constrained random phase approximation in the Wannier basis~\cite{rpa1,rpa2}, as implemented in \texttt{VASP}~\cite{rpa3}. 

\begin{figure}
\begin{center}
\includegraphics[width=0.49\textwidth]{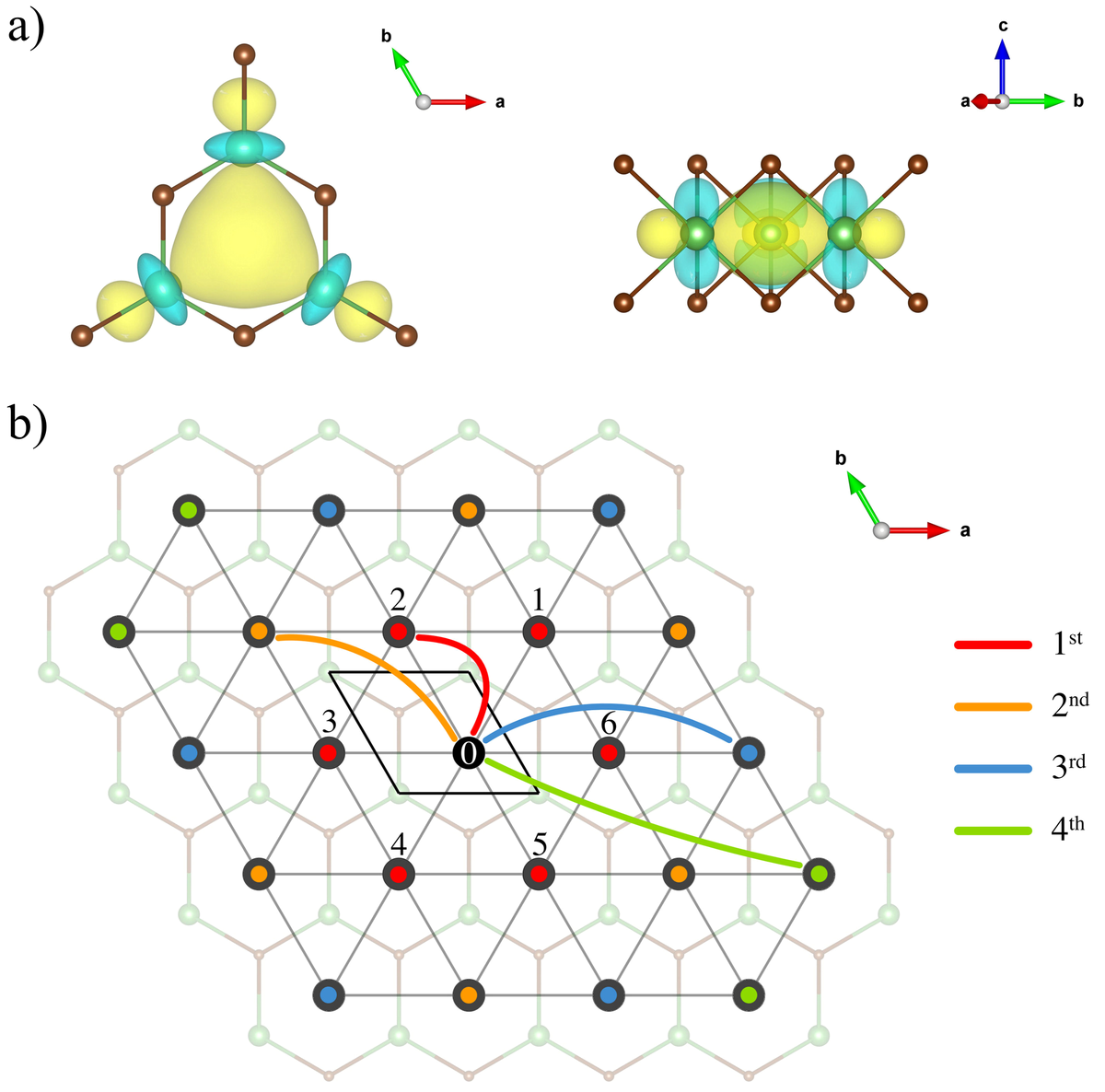}
\end{center}
\caption{a) Wannier function representing the anionic electron in LaBr$_{2}$, as obtained from LDA calculations; b) Schematic view of the triangular lattice formed by the anionic electrons.}
\label{fig:wan1}
\end{figure}

\begin{figure}
\begin{center}
\includegraphics[width=0.48\textwidth]{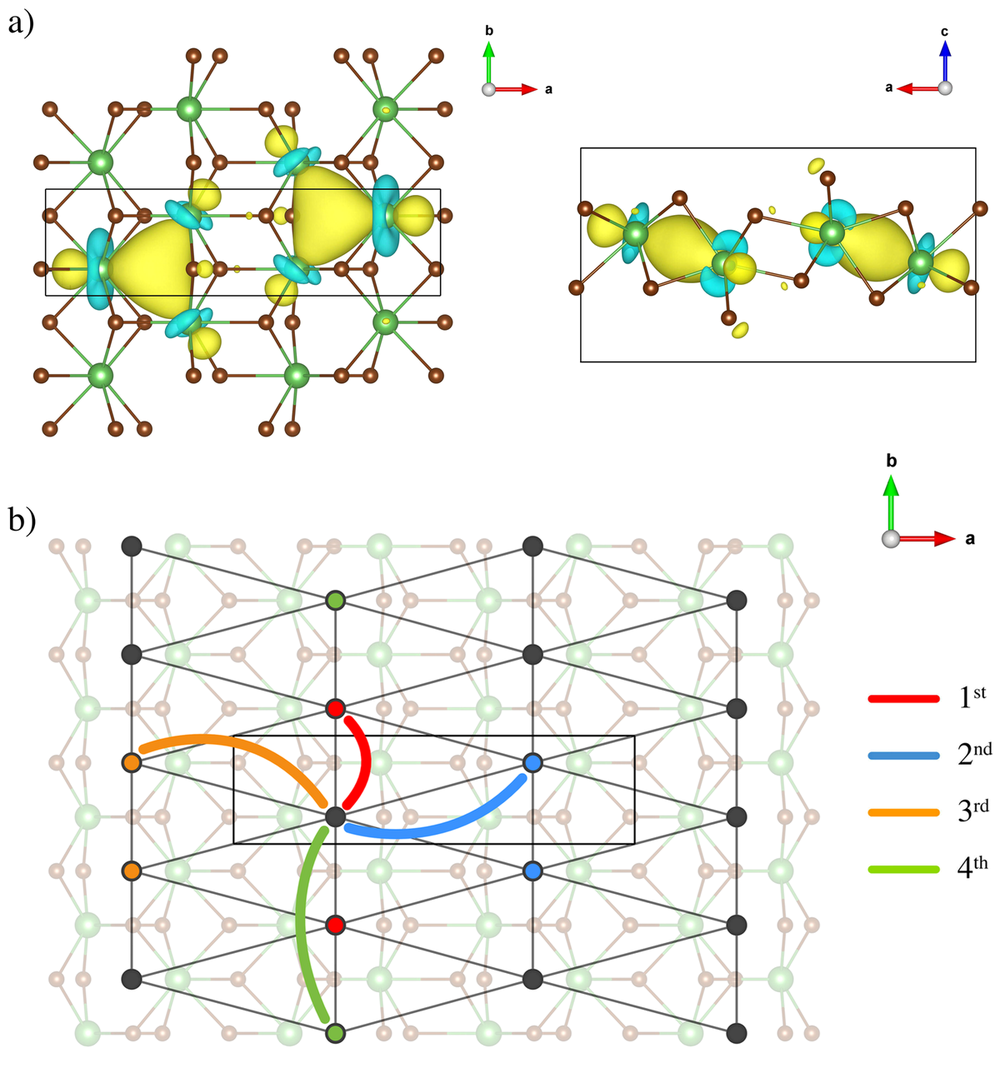}
\end{center}
\caption{a) Wannier function representing the anionic electrons in La$_{2}$Br$_{5}$, as obtained from LDA calculations; b) Schematic view of the lattice formed by the anionic electrons.}
\label{fig:wan2}
\end{figure}

\par As shown in figures~\ref{fig:wan1}a and \ref{fig:wan2}a, the Wannier functions representing the anionic bands in LaBr$_{2}$ and La$_{2}$Br$_{5}$ are located at the center of hexagons extending their tails to the neighbouring La sites. According to our calculations carried out with different projectors as an initial guess for the wannierisation routine~\cite{supp}, the resulting Wannier orbital not only has a minimal possible spatial spread (7.75~\AA$^{2}$ in LaBr$_{2}$ and 9.56~\AA$^{2}$ in La$_{2}$Br$_{5}$), but also maximises its on-site Coulomb interaction, thus justifying this choice as a proper basis for the low-energy effective model~\cite{coulomb}. However, despite being well localised at interstitial cavities, they reveal a very extended nature commonly inherent to delocalised orbitals. Such a duality of the anionic states will play an ultimate role in defining magnetic properties of the considered electrides.

\par The resulting model parameters for LaBr$_{2}$ and La$_{2}$Br$_{5}$ are presented in Tables~\ref{tab:1} and \ref{tab:2}, respectively. Following the $D_{3h}$ symmetry of monolayer LaBr$_{2}$, $\hat{t}_{ij}=\parallel \!t_{ij}^{\sigma\sigma'}\!\!\parallel$ between first nearest neighbours can be written $\hat{t}_{ij}=t_{}\hat{\sigma}_{0}+(-1)^{j}\mathtt{i}t'_{}\hat{\sigma}_{z}$, as shown in figure~\ref{fig:wan1}b, where $\mathtt{i}$ is the imaginary unit, $\sigma_{0}$ and $\sigma_{n}$ are the unitary matrix and the $n^{\mathrm{th}}$ component Pauli matrix, respectively (the same form applies to third and fourth nearest neighbours, although their contribution is negligibly small). In monolayer La$_{2}$Br$_{5}$, given the $C_{2h}$ symmetry with the principal monoclinic $\mathbf{b}$ axis and two anionic electrons in the unit cell, the hopping parameters from site $i=1,2$ obey the following form $\hat{t}_{ii\pm\mathbf{b}}=t_{}\hat{\sigma}_{0}\pm(-1)^{i}(\mathtt{i}t'_{}\hat{\sigma}_{z}+\mathtt{i}t''_{}\hat{\sigma}_{x})$ for nearest neighbours in the $\pm\mathbf{b}$ directions, and $\hat{t}_{ij}=t_{}\hat{\sigma}_{0}$ for other neighbours, as shown in figure~\ref{fig:wan2}b. Since the anionic electrons are loosely bound, the hopping parameters are found to be relatively small in both systems.

\begin{table}[t!]
\caption{Parameters of the one-orbital extended Hubbard model for LaBr$_{2}$. Nearest neighbours with the distance $d$ are shown in figure~\ref{fig:wan1}b. The on-site Coulomb interaction $U$ is 1.54~eV. }
\begin{center}
\begin{tabular}{c|cccc}
\hline
\hline
& 1$^{st}$ & 2$^{nd}$ & 3$^{rd}$ & 4$^{th}$  \\
$d$ (\AA) &  4.099 & 7.099 & 8.198 & 10.844 \\
\hline
$t_{}$ (meV) & 14.1 & 64.9 & $-14.7$ & $-2.7$ \\ 
$t'_{}$ (meV) & 5.3 & 0.0 & 0.2 & 0.3 \\ 
$V$ (eV) & 0.72 & 0.61 & 0.43 & 0.26 \\
$J^{F}$ (meV) & 20.3 & $4.4$ & $-$ & $-$ \\
\hline
\hline
\end{tabular}
\end{center}
\label{tab:1}
\end{table}

\begin{table}[b!]
\caption{Parameters of the one-orbital extended Hubbard model for La$_{2}$Br$_{5}$. Nearest neighbours with the distance $d$ are given according to figure~\ref{fig:wan2}b. The on-site Coulomb interaction $U$ is 2.04 eV. }
\begin{center}
\begin{tabular}{c|cccc}
\hline
\hline
 & 1$^{st}$ &  2$^{nd}$ & 3$^{rd}$ & 4$^{th}$   \\
\!\!$d$ (\AA) & 4.301 & 8.127 & 8.379 & 8.601 \\
\hline 
\!\!$t$ (meV) & $-18.8$ & $-1.4$ & 12.0 & $-44.0$  \\
\!\!$t'$ (meV) &  $-4.6$ & 0 & 0 & 1.0  \\
\!\!$t''$ (meV) & 0.8 & 0 & 0 & 1.4   \\
\!\!$V$ (eV)  & 1.18 & 0.83 & 0.65 & 0.73 \\
\!\!$J^{F}$ (meV) & 25.0 & $-$ & $-$ & $6.7$  \\
\hline
\hline
\end{tabular}
\end{center}
\label{tab:2}
\end{table}

\par It is notable that in both system the anionic states located at the Fermi level are well separated from the rest of the electronic spectrum. As a result, they end up being weakly screened by the atomic states that results in reasonably large on-site $U$ and intersite~$V$. Associated with the extended nature of anionic electrons, Coulomb interactions between neighbouring orbitals decrease slowly with distance slightly reducing the effect of on-site $U$. Moreover, due to their large spatial extension, the overlap of neighbouring anionic orbitals leads to a sizeable direct exchange interaction. According to our calculations \cite{supp}, the bare $J^{F}$ decays faster over the distance than $V$, and bearing in mind its partial screening by the rest of the system it takes a large value only between first nearest neighbours. Given the loosely bound character of the anionic electrons, this interaction in turn will dominate over kinetic superexchange stabilising a ferromagnetic order in these electride systems. 

\begin{figure}
\begin{center}
\includegraphics[width=0.48\textwidth]{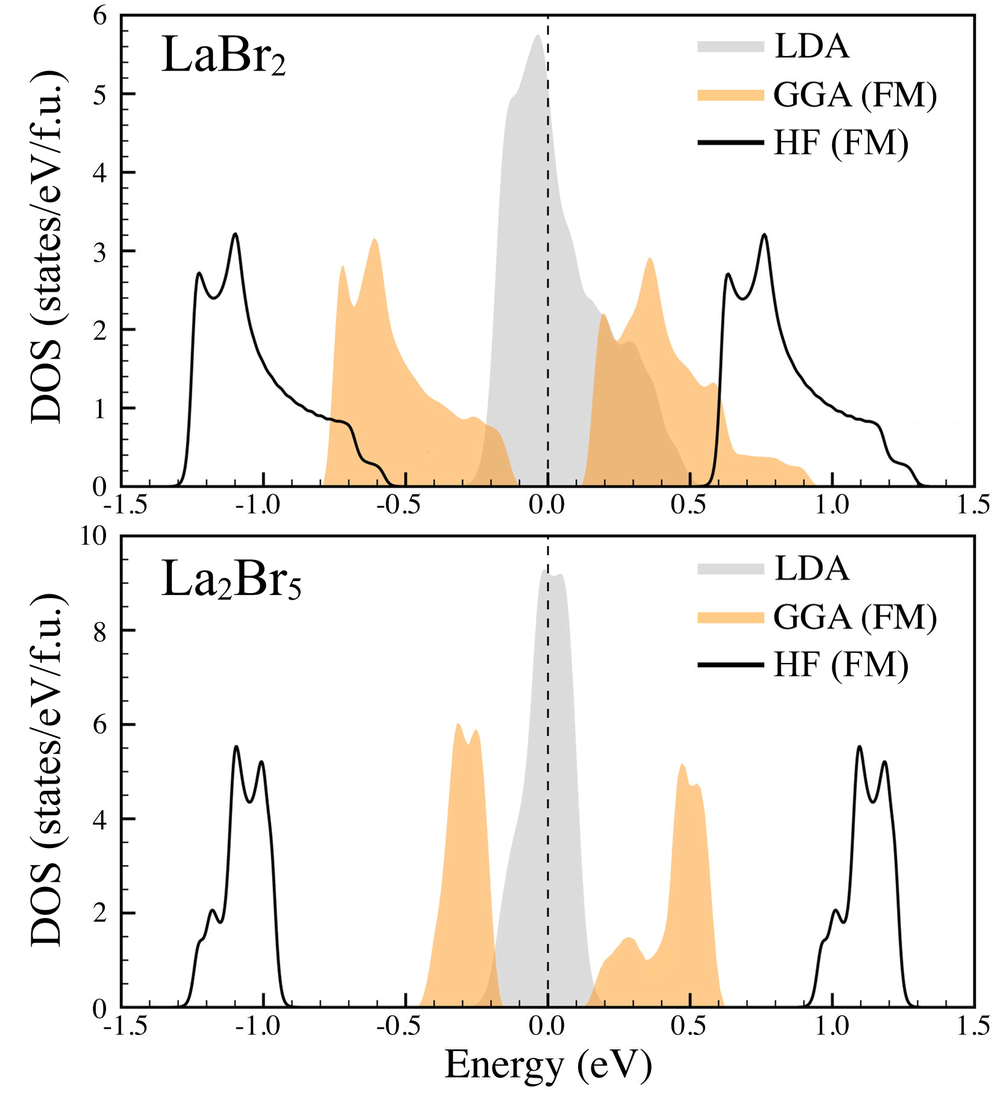}
\end{center}
\caption{Densities of states in LaBr$_{2}$ (a) and La$_{2}$Br$_{5}$ (b) obtained in Hartree-Fock calculations for the ferromagnetic collinear configurations.}
\label{fig:dos}
\end{figure}

\par Taking into account the limit of large on-site Coulomb interactions $U\gg t$, the constructed Hubbard models can be solved within the Hartree-Fock approximation assuming a collinear magnetic configuration~\cite{hf1,hf2}: 
\begin{equation}
\Big(\hat{t}_{\mathbf{k}}+\hat{\mathcal{V}}^{\sigma}_{\mathrm{HF}}\big)|\phi_{\mathbf{k}}^{\sigma}\rangle = \varepsilon^{\sigma}_{\mathbf{k}}|\phi^{\sigma}_{\mathbf{k}}\rangle
\end{equation}
\noindent where $\hat{t}_{\mathbf{k}}$ is the Fourier transform of hopping parameters, $\varepsilon^{\sigma}_{\mathbf{k}}$ and $|\phi^{\sigma}_{\mathbf{k}}\rangle$ are eigenvalues and eigenvectors in the Wannier basis, respectively, and the Hartree-Fock potential at site $i$ is defined as:
\begin{equation}
\mathcal{V}_{\mathrm{HF},i}^{\sigma}=U n^{\bar{\sigma}}_{i} +\sum_{j}V_{ij}n^{}_{j}+\sum_{j}J_{ij}n_{j}^{\sigma},
\end{equation}
\noindent where $n^{\sigma}_{i}=\sum_{\mathbf{k}}|\phi_{\mathbf{k}}^{\sigma}\rangle\langle\phi_{\mathbf{k}}^{\sigma}|$, $n^{}_{j}=n^{\uparrow}_{j}+n^{\downarrow}_{j}$, $\bar{\sigma}$ takes the opposite direction of spin $\sigma$, and the intersite density terms are neglected on account of their smallness. A self-consistent solution of the Hartree-Fock equations is obtained with respect to the total energy:
\begin{equation}
E_{\mathrm{HF}}=\sum_{\mathbf{k}\sigma}\varepsilon_{k}^{\sigma}-\frac{1}{2}\sum_{i\sigma} \mathcal{V}^{\sigma}_{\mathrm{HF},i}n^{\sigma}_{i}.
\end{equation}
\noindent The resulting densities of states calculated for the ferromagnetic configurations are shown in figure~\ref{fig:dos}, revealing sizeable band gaps of $\sim$1.0 eV and $\sim$1.8 eV in LaBr$_{2}$ and La$_{2}$Br$_{5}$, respectively, and clearly indicating a Mott-insulating regime for the anionic electrons. Consequently, one can see that conventional LDA and GGA calculations largely underestimate the insulating gap in these systems. 
\par \emph{Magnetic ground state}. In the limit $U>V\gg t$, the Hubbard model can be mapped onto the spin model by using the theory of superexchange~\cite{super1,super2,super3,super4}:
\begin{equation}
\mathcal{H}_{S}=\sum_{\langle ij \rangle} J_{ij}\boldsymbol{S}_{i}\cdot\boldsymbol{S}_{j} + \boldsymbol{D}_{ij}\cdot(\boldsymbol{S}_{i}\times\boldsymbol{S}_{j})+\boldsymbol{S}_{i}\Gamma_{ij}\boldsymbol{S}_{j},
\end{equation}
\noindent with 
\begin{gather*}
J_{ij}=\frac{2}{U_{i}-V_{ij}}\mathrm{Tr}[\hat{t}_{ij}\hat{t}_{ji}]-2J_{ij}^{F}, \\
\boldsymbol{D}_{ij}=\frac{\mathtt{i}}{U_{i}-V_{ij}}\big(\mathrm{Tr}[\hat{t}_{ij}]\mathrm{Tr}[\hat{t}_{ji}\hat{\boldsymbol{\sigma}}]- i \leftrightarrow j \big), \\
\Gamma_{ij}=\frac{1}{U_{i}-V_{ij}}\big(\mathrm{Tr}[\hat{t}_{ij}\hat{\boldsymbol{\sigma}}]\otimes\mathrm{Tr}[\hat{t}_{ji}\hat{\boldsymbol{\sigma}}] + i \leftrightarrow j \big), 
\end{gather*}
\noindent where the first term is the isotropic Heisenberg exchange written as a sum of antiferromagnetic superexchange interactions competing with direct ferromagnetic exchange, and the last two terms stand for the antisymmetric Dzyaloshinskii-Moriya and symmetric anisotropic exchange interactions, respectively. Given the $S=1/2$ spin systems, we do not consider single-ion anisotropy. 

\begin{figure}[b]
\begin{center}
\includegraphics[width=0.50\textwidth]{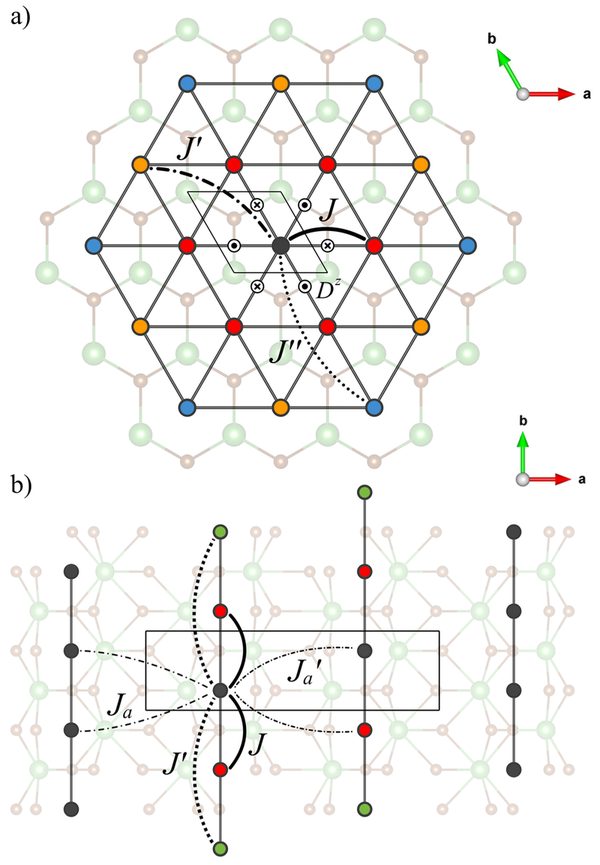}
\end{center}
\caption{Schematics of the spin models in a) LaBr$_{2}$ and b) La$_{2}$Br$_{5}$.}
\label{fig:spin}
\end{figure}

\begin{table}[b]
\caption{Isotropic exchange parameters (in meV) between anionic states in LaBr$_{2}$ and La$_{2}$Br$_{5}$. The kinetic and direct contributions are given by the first second terms in the definition of $J_{ij}$, respectively.}
\begin{center}
\begin{tabular}{c|rrr}
\hline
\hline
LaBr$_{2}$ & $J$ & $J'$ & $J''$  \\
\hline
kinetic & 1.0 & 18.1 & 0.8 \\ 
direct & $-40.6$ & $-8.8$ & 0 \\ 
total & $-39.6$ & 9.3 & 0.8 \\
\hline
\hline
\end{tabular}\\
\begin{tabular}{c|rrrr}
\hline
\hline
La$_{2}$Br$_{5}$ & $J$ & $J'$ & $J_{a}$ & $J_{a}'$  \\
\hline
kinetic & 1.6 & 5.9 & 0.4 & $\sim$0\\ 
direct & $-50.0$ & $-13.4$ & 0  & 0\\ 
total & $-48.4$ & $-7.5$ & 0.4 & $~\sim$0 \\
\hline
\hline
\end{tabular}
\end{center}
\label{tab:3}
\end{table}

\par The resulting spin models derived for LaBr$_{2}$ and La$_{2}$Br$_{5}$ are sketched in figure~\ref{fig:spin}. From what has been said, it follows that strong direct exchange mediated by the large overlap between neighbouring anionic electrons can outweigh its kinetic superexchange counterpart, as clearly seen in Table~\ref{tab:3}. As a result, the anionic states in LaBr$_{2}$ form a triangular lattice with strong ferromagnetic and weak antiferromagnetic couplings between first and second nearest neighbours, respectively. In turn, intersite anisotropy in LaBr$_{2}$ is given by alternating Dzyaloshinskii-Moriya $D^{z}=\pm0.7$~meV and symmetric anisotropic $\Gamma^{zz}=$0.3~meV interactions between first nearest neighbours, whose components are in agreement with Moriya rules (the in-plane mirror reflection of the $D_{3h}$ symmetry passing through all magnetic sites gives the direction of $\boldsymbol{D}$ perpendicular to this plane)~\cite{super2}. In a similar manner, strong direct exchange in La$_{2}$Br$_{5}$ dominates over kinetic superexchange forming ferromagnetic chains along the $\mathbf{b}$ axis, which are weakly antiferromagnetically coupled in the $\mathbf{a}$ direction. Intersite anisotropic interactions between intrachain nearest neighbours are given by $\boldsymbol{D}=(\mp D^{x},0,\pm D^{z})$, with $D^{x}=$0.1 meV, $D^{z}=$0.8 meV in the $\pm\mathbf{b}$ direction, and $\Gamma^{zz}=$0.2~meV, also satisfying Moriya rules (the $C_{2h}$ mirror symmetry in the $ac$ plane passes though the middle of the bond between neighbouring spins along the $\mathbf{b}$ axis and gives $\boldsymbol{D}$ lying in the $ac$ plane).

\par It is well known that anisotropy favouring the in-plane orientation of magnetic moments does not produce a gap in the magnon spectra of 2D spin systems. Therefore, the formation of any long-range order in LaBr$_{2}$ is prohibited by Mermin-Wagner theorem~\cite{mermin}. However, several possibilities can still be employed to stabilise a magnetic order, such as considering magnetic dipole-dipole interactions~\cite{bruno}, inducing a uniaxial strain, or applying a small external magnetic field. In contrast, having a lower symmetry the intersite anisotropic interactions between anionic electrons in monolayer La$_{2}$Br$_{5}$ align the ferromagnetic chains perpendicular to the $ac$ plane, thus stabilising a long-range order along the $\mathbf{b}$ axis.

\par Finally, it is worth comparing our electrenes with other magnetic materials featuring similar microscopic properties. Strictly speaking, the considered mechanism stabilising ferromagnetism in 2D systems is not new. For example, an important role of direct ferromagnetic exchange was previously reported in functionalised graphene derivatives C$_{2}$H and C$_{2}$F, where the same approach was used to construct effective electronic models~\cite{hf2}. The Wannier functions associated with the states at the Fermi level were shown to have large spreads of 3.10~\AA$^{2}$ and 2.66~\AA$^{2}$ in C$_{2}$H and C$_{2}$F, respectively, where given shorter distances of $\sim$2.5~\AA~between neighbouring orbitals, their overlap results in strong direct exchange ranging from 0.02~eV to 0.10~eV and dominating over antiferromagnetic kinetic contributions. Despite having larger distances, the overlap between neighbouring anionic states in LaBr$_{2}$ and La$_{2}$Br$_{5}$ decreases slowly in space mediating large intersite Coulomb interactions beyond first nearest neighbours, whereas their direct exchange decays faster and is also essentially short-range as in the C$_{2}$H and C$_{2}$F systems. Thus, regardless of the ``orbital" character of magnetic states, the same origin of intrinsic magnetism stemming entirely from the band structure can be common to many systems, whose localised orbitals display an extended nature.

\par \emph{Conclusions}. Having combined first-principles calculations and effective electronic models, we studied intrinsic magnetism in two-dimensional electrides LaBr$_{2}$ and La$_{2}$Br$_{5}$. Being confined at zero-dimensional cavities in the crystal structure, the anionic electrons in these systems are well localised but largely extended featuring such a dual nature with no atomic orbitals involved. On the one hand, the anionic states are subject to strong on-site Coulomb interaction and reveal a Mott-insulating state with localised magnetic moments. On the other hand, owing to the large spatial extension their overlap gives rise to a dominating direct ferromagnetic exchange between neighbouring electrons that determines magnetic properties of the systems in question. While it was shown that any long-range magnetic order is prohibited in LaBr$_{2}$ by Mermin-Wagner theorem, the ferromagnetic chains oriented along the monoclinic $\mathbf{b}$ axis can be stabilised in La$_{2}$Br$_{5}$. At last, we showed that conventional first-principles calculations can be inadequate in describing magnetism of electride materials with localised anionic states. Their striking difference with ubiquitous $sp$- and $d$-electron systems is that the magnetic moments are formed at interstitial regions rather than at atomic orbitals making them a unique example among strongly correlated materials.

\par \emph{Acknowledgements}. The authors would like to thank Dr.~Lee~A.~Burton for numerous stimulating discussions.


\end{document}


\title{Supplemental Material:\\ Localised Magnetism in 2D Electrides}

\author{Danis~I.~Badrtdinov}
\affiliation{Department of Theoretical Physics and Applied Mathematics, Ural Federal University, Mira St. 19, 620002 Yekaterinburg, Russia}
\author{Sergey~A.~Nikolaev}
\email{nikolaev.s.aa@m.titech.ac.jp}
\affiliation{Institute of Innovative Research, Tokyo Institute of Technology, 4259 Nagatsuta, Midori, Yokohama 226-8503, Japan}
\affiliation{Department of Theoretical Physics and Applied Mathematics, Ural Federal University, Mira St. 19, 620002 Yekaterinburg, Russia}

\maketitle

\section{First-principle calculations}
\par Electronic structure calculations were performed within local density approximation (LDA)~\cite{lda} and generalized gradient approximation (GGA)~\cite{pbe}, as implemented in the projector augmented wave based Vienna ab-initio simulation package \texttt{VASP}~\cite{vasp}, plane-wave based \texttt{Quantum Espresso} package (\texttt{QE})~\cite{qe}, and all-electron full-potential linearised augmented plane wave \texttt{ELK} code~\cite{elk}. For calculations in \texttt{QE} we used norm-conserving pseudopotentials~\cite{norm}. The plane-wave cutoff was set to 500 eV and 1360 eV for \texttt{VASP} and \texttt{QE}, respectively, and the convergence criteria for the total energy calculations was set to 10$^{-9}$ eV. Calculations in \texttt{ELK} were carried out with $R_{\mathrm{MT}}*|G + k|_{\mathrm{max}}$ = 7.0. The Brillouin zone in all calculations was sampled by $30\times30\times1$ and $10\times20\times1$ Monkhorst-Pack grids~\cite{mp} for LaBr$_2$ and La$_2$Br$_2$, respectively.  Crystallographic data of the bulk systems was adopted from the Materials project database~\cite{materials_project}, and separate monolayers were taken with a vacuum space of 18 \AA~between unit cell replicas in the vertical $z$ direction. Atomic positions of the constructed monolayers were relaxed until the residual forces were less than 0.001 eV/\AA, and the optimised crystal structures are given in Table~\ref{tab:Structural_parameters}.

\par
\begin{figure}[!h]
\includegraphics[width=1.0\textwidth]{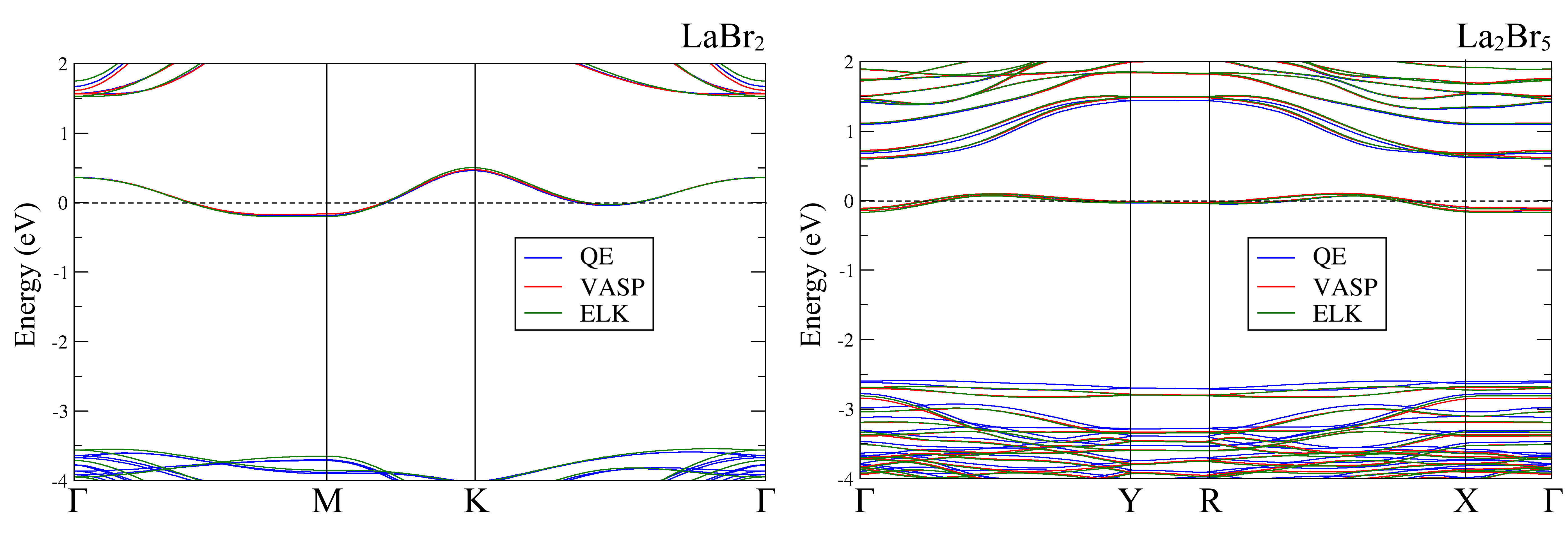}
\caption{Comparison of the band structures of LaBr$_2$ (left) and La$_2$Br$_5$ (right) obtained from GGA calculations in \texttt{QE},  \texttt{ELK}, and \texttt{VASP} codes. ``Ghost'' states corresponding to the La $f$ electrons found in \texttt{VASP} were shifted to the highly unoccupied states by applying the on-site $U$=10 eV. The high-symmetry $k$-points are M=$(0,\frac{1}{2},0)$, K=$(\frac{1}{3},\frac{1}{3},0)$ for LaBr$_2$,  and Y=$(0,\frac{1}{2},0)$, R=$(\frac{1}{2},\frac{1}{2},0)$, X=$(\frac{1}{2},0,0)$ for La$_2$Br$_5$.}
\label{fig:BANDS}
\end{figure}

 In \texttt{VASP} calculations, we have found the ``ghost" states located slightly above the Fermi level and corresponding to the La $f$-electrons. These bands can be effectively eliminated by pushing them further to the highly unoccupied states with the on-site Coulomb parameter $U$ ($U \geq 10$ eV) turned on at the La sites, that leads to a good agreement of the electronic structure with other codes. Their comparison is illustrated in figure~\ref{fig:BANDS}. \vspace{0.5cm}

\begin{table}[!h]
\centering
\caption{Crystal structures of monolayer  LaBr$_2$ and La$_2$Br$_5$, as optimised within non spin-polarised GGA calculations in \texttt{ELK}. }
\begin{ruledtabular}
\begin{tabular}{ccccccc}
\multicolumn{3}{c}{LaBr$_2$}  &  &  \multicolumn{3}{c}{La$_2$Br$_5$} \\
 \cline{1-3} \cline{5-7}  
\multicolumn{3}{c}{ $a$ = $b$ =  4.0988 \AA, $c$ = 18 \AA; }  &   &  \multicolumn{3}{c}{$a$ = 15.9357 \AA, $b$ = 4.3000 \AA, $c$ =  18 \AA; } \\
\multicolumn{3}{c}{ $\alpha = \beta = 90^\circ$, $\gamma = 120^\circ$ }  &  &  \multicolumn{3}{c}{$\alpha = \beta = \gamma = 90^\circ$} \\
\multicolumn{3}{c}{ Space group is $P6_{3}/mmc$ (No. 194) }  &   & \multicolumn{3}{c}{ Space group is $P2_1/m$ (No. 11)}  \\ 
 \cline{1-3} \cline{5-7}  
 &&&&&& \\
La  &  $2b$  & ($\frac{1}{3}$, $\frac{2}{6}$, 0.2571)   &  &   La1     &   2$e$    &   (0.8608, $\frac{1}{4}$, 0.5321)  \\
Br   &  $4f$  &(0, 0, $\frac{1}{4}$)    &  &   La2    &   2$e$    &   (0.3636, $\frac{1}{4}$, 0.5347)   \\
       &    &                                                               &  &   Br1     &   2$e$    &    (0.2506, $\frac{1}{4}$, 0.3878)   \\
       &    &                                                               &  &    Br2    &  2$e$    &    (0.0111, $\frac{1}{4}$, 0.4114)  \\
       &    &                                                               &  &    Br3    &  2$e$    &    (0.5585, $\frac{1}{4}$, 0.5727)   \\
       &    &                                                               & &    Br4    &  2$e$     &    (0.6251, $\frac{1}{4}$, 0.3437)   \\
       &    &                                                               &  &    Br5    &  2$e$    &    (0.1774, $\frac{1}{4}$, 0.5895)   \\         
&&&&&& \\
\end{tabular}
\end{ruledtabular}
\label{tab:Structural_parameters}
\end {table}

\par
\begin{figure}[!h]
\includegraphics[width=1.0\textwidth]{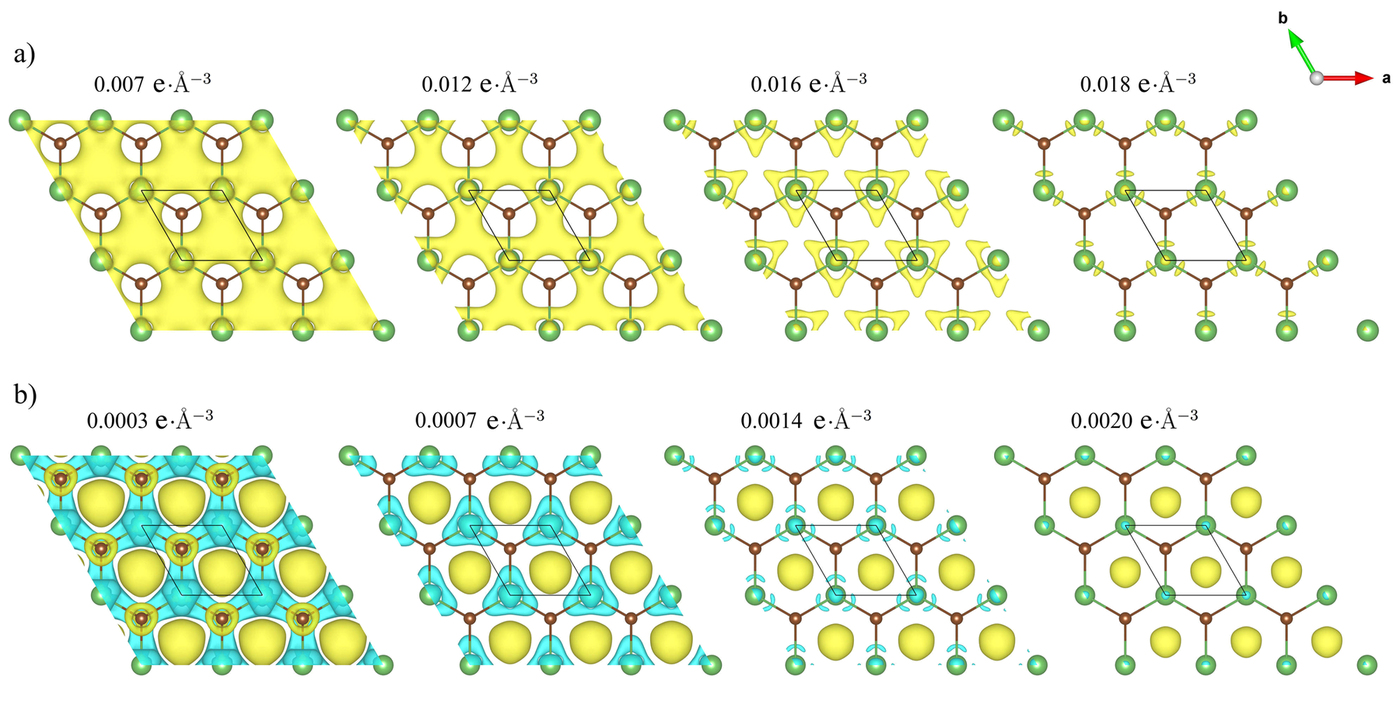}
\caption{Isosurfaces of the charge (a) and magnetisation (b) densities corresponding to the anionic band in the ferromagnetic state of LaBr$_{2}$, as obtained within GGA calculations in \texttt{VASP}. The unit cells are shown by black boxes.}
\label{fig:Anionic_band}
\end{figure}

\begin{figure}[!h]
\includegraphics[width=1.0\textwidth]{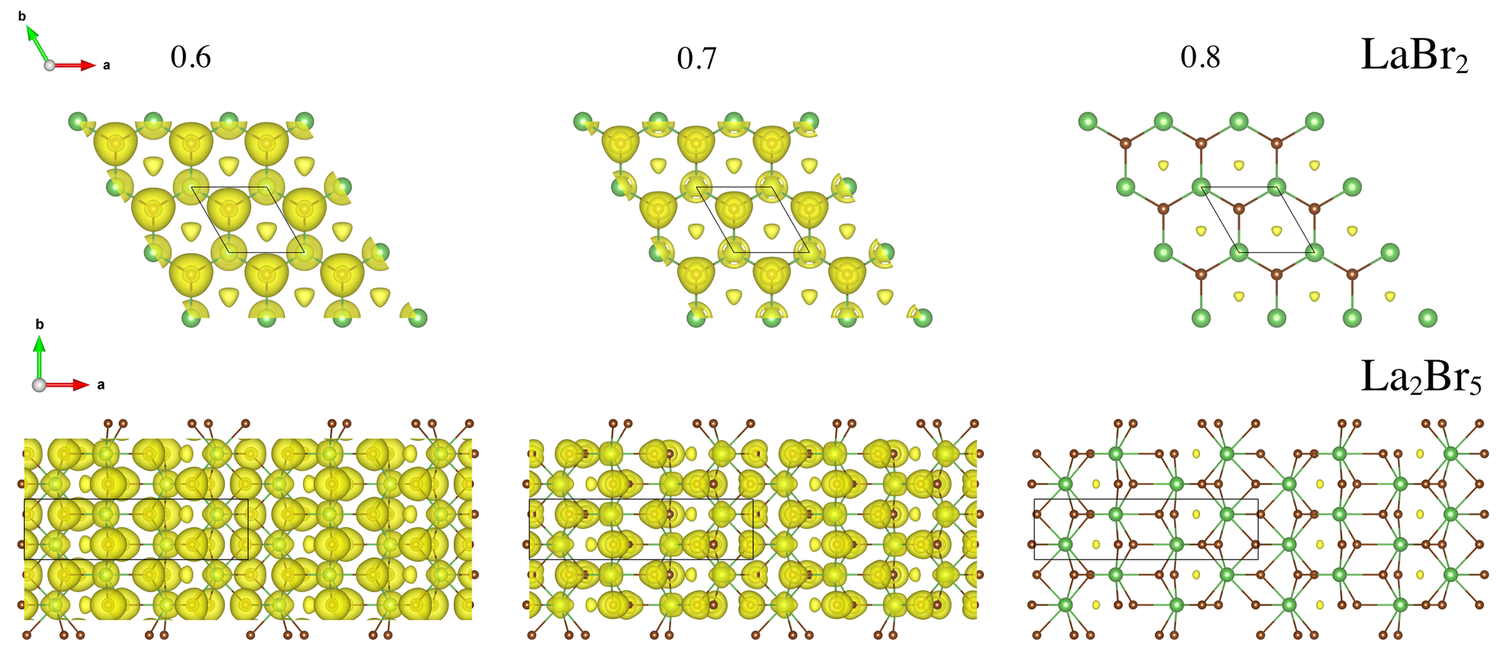}
\caption{Isosurfaces of the electron localisation function (ELF) as obtained for the ferromagnetic state of LaBr$_{2}$ and La$_{2}$Br$_{5}$ within GGA calculations in \texttt{VASP}. The unit cells are shown by black boxes.}
\label{fig:elf}
\end{figure}

\begin{figure}[!h]
\includegraphics[width=1.0\textwidth]{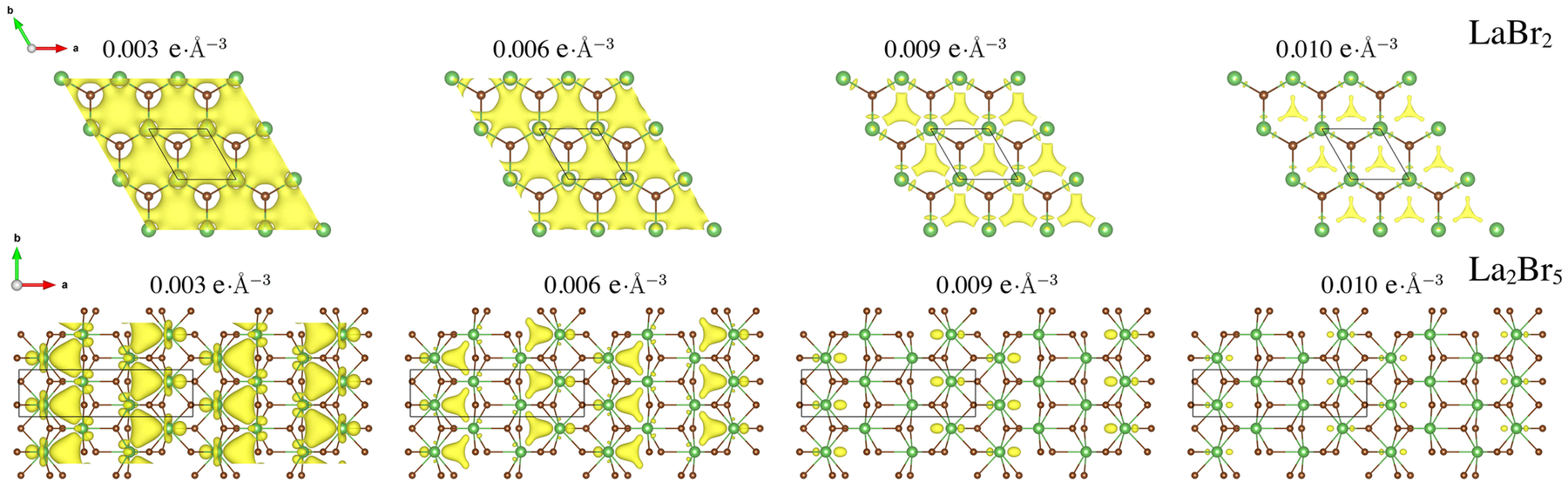}
\caption{Isosurfaces of the total magnetisation densities as obtained for the ferromagnetic state in LaBr$_{2}$ and La$_{2}$Br$_{5}$ within GGA calculations in \texttt{VASP}. The unit cells are shown by black boxes. }
\label{fig:magtot}
\end{figure}

 According to our first-principles calculations, the electronic states at the Fermi level correspond to the excess electrons localised at the center of hexagons. For the sake of clarity, several isosurfaces of the charge and magnetisation densities associated with the anionic bands in the ferromagnetic state of LaBr$_{2}$ are shown in figure~\ref{fig:Anionic_band}. For illustration purposes, we also present the total magnetisation density and electron localisation function (ELF) in figures~\ref{fig:elf} and \ref{fig:magtot}, respectively, calculated for the ferromagnetic configurations in LaBr$_2$  and La$_{2}$Br$_{5}$. Our calculations clearly indicate that the anionic electrons are confined in hexagonal cavities.

\section{Wannier functions}
\par
\begin{figure}[!h]
\includegraphics[width=0.70\textwidth]{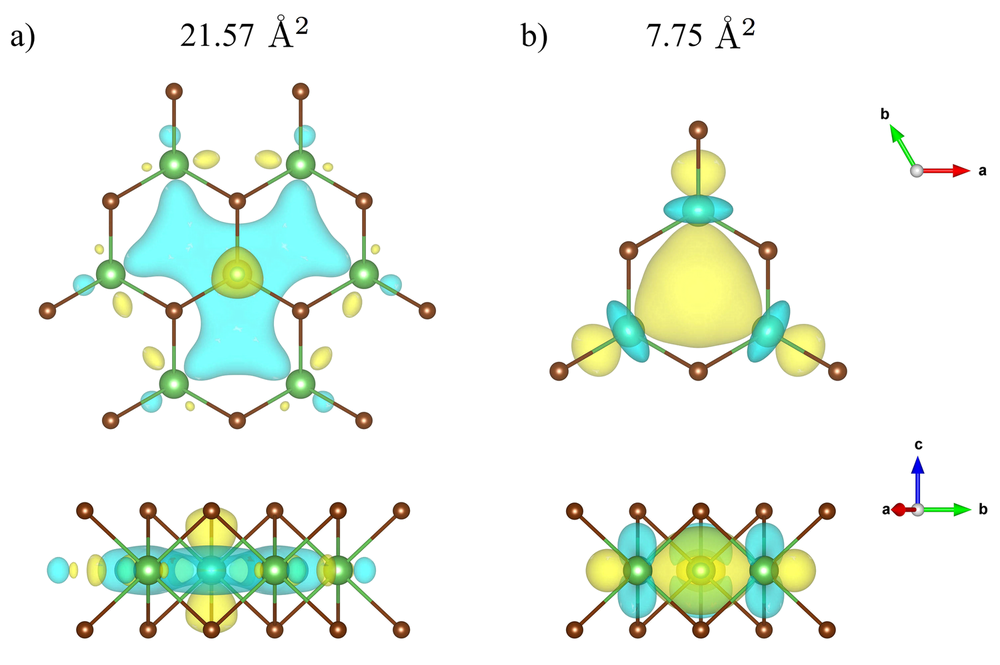}
\caption{Maximally localised Wannier functions representing the anionic electron in LaBr$_{2}$ as obtained from LDA calculations in \texttt{QE} by projecting the band at the Fermi level onto: a) $d_{z^{2}}$-orbital centered at the La atom; b) $s$-orbital centered at the hexagon. The numbers indicated above correspond to the calculated spread $\Omega=\langle r \rangle ^{2} - \langle r^{2} \rangle$.}
\label{fig:wan}
\end{figure}

 The basis of Wannier functions was constructed from electronic structure calculations by using the procedure of maximal localisation as implemented in the \texttt{wannier90} package~\cite{wan}. Numerically, one can employ different initial orbitals to project the anionic band. As illustrated in figure~\ref{fig:wan}, the resulting shape of the Wannier function corresponding to the anionic band in LaBr$_{2}$ strongly depends on the starting guess. Regardless of having the same dispersion, the Wannier functions constructed from the $s$-orbitals located at the center of hexagons reveal a smaller spread (defined as $\Omega=\langle r \rangle ^{2} - \langle r^{2} \rangle$) compared to the ones obtained starting from the $d$-orbitals centred at the La atoms, justifying the former as a proper basis with maximal localisation. Moreover, the Wannier functions centered at the hexagons maximise the on-site Coulomb interactions, which can be regarded as another criterion for constructing effective orbitals~\cite{coulomb}. Using cRPA calculations, the obtained value of the on-site Coulomb interactions $U$ is 0.47 eV for the La centered orbitals, compared to 1.54 eV for the orbitals located at the hexagon.  
\par
\begin{figure}[!h]
\includegraphics[width=1.0\textwidth]{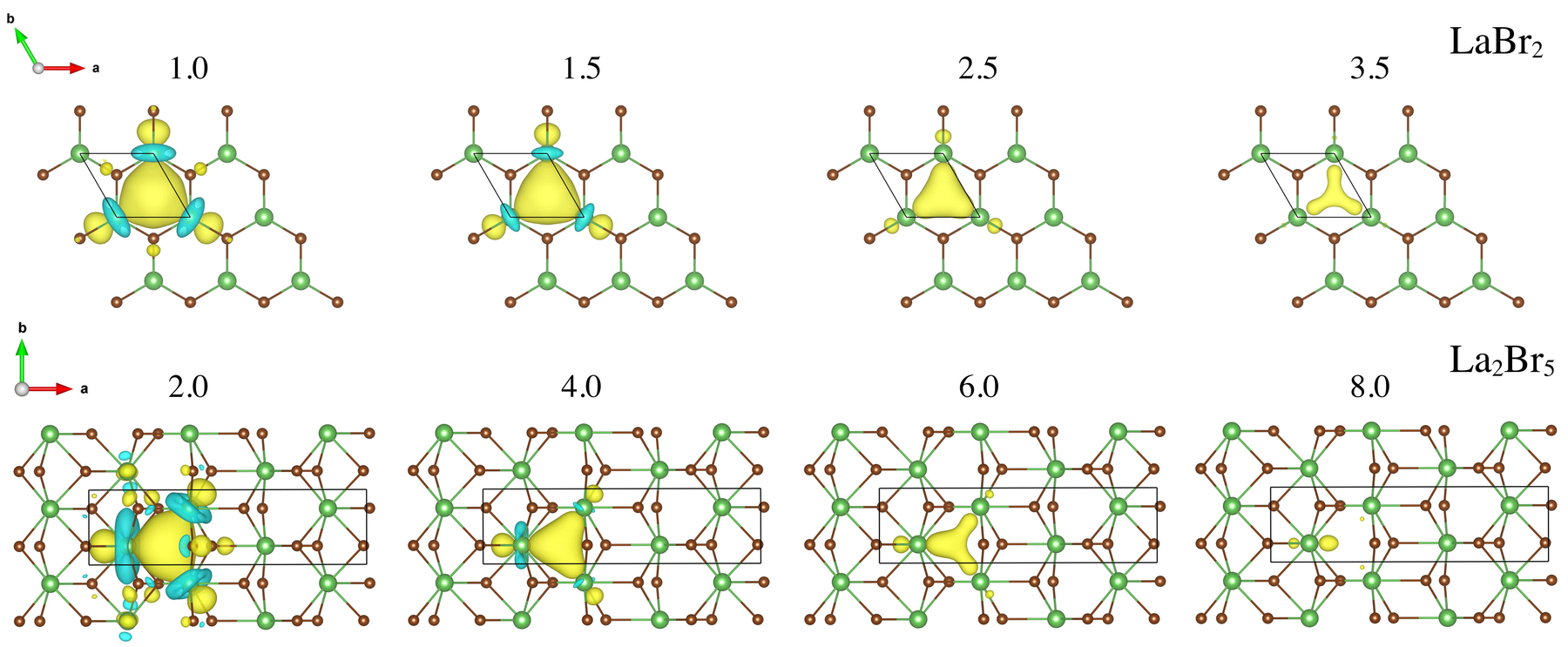}
\caption{Isosurfaces of the maximally localised Wannier functions as obtained from LDA calculations in \texttt{QE} for (c) LaBr$_{2}$ and (d) La$_{2}$Br$_{5}$. The unit cells are shown by black boxes.}
\label{fig:wan_all}
\end{figure}

 Isosurfaces of the Wannier functions corresponding to the anionic states in LaBr$_{2}$ and La$_{2}$Br$_{5}$ are shown in figure~\ref{fig:wan_all}. 

\section{Extended Hubbard model}
\par We consider an extended Hubbard-type model in the basis of Wannier functions representing the anionic states:
\begin{equation}
\begin{aligned}
\mathcal{H}&=\mathcal{H}_{\mathrm{kin}}+\mathcal{H}_{U}+\mathcal{H}_{V}+\mathcal{H}_{J}\\
&=\sum_{ij,\sigma\sigma'}t_{ij}^{\sigma\sigma'}a^{\dagger\sigma}_{i}a^{\sigma'}_{j} + \frac{1}{2}\sum_{i,\sigma\sigma'}U_{i}a^{\dagger\sigma}_{i}a^{\dagger\sigma'}_{i}a^{\phantom{\dagger}\!\!\sigma'}_{i}a^{\phantom{\dagger}\!\!\sigma}_{i} + \frac{1}{2}\sum_{ij,\sigma\sigma'}V_{ij}a^{\dagger\sigma}_{i}a^{\dagger\sigma'}_{j}a^{\phantom{\dagger}\!\!\sigma'}_{j}a^{\phantom{\dagger}\!\!\sigma}_{i}  + \frac{1}{2}\sum_{ij,\sigma\sigma'}J_{ij}^{F}a^{\dagger\sigma}_{i}a^{\dagger\sigma'}_{j}a^{\phantom{\dagger}\!\!\sigma'}_{i}a^{\phantom{\dagger}\!\!\sigma}_{j},
\end{aligned}
\end{equation}
\noindent where $a^{\dagger\sigma}_{i}$ ($a^{\phantom{\dagger}\!\!\sigma}_{i}$) creates (annihilates) an electron with spin $\sigma$ at  the anionic site $i$, and $t_{ij}^{\sigma\sigma'}$ is the hopping parameter including spin-orbit coupling between neighbouring sites. Here, $U_{i}$ is the on-site Coulomb interaction; $V_{ij}$ and $J_{ij}$ stand for the intersite Coulomb and exchange interactions, respectively. Matrix elements of the hopping parameters were obtained by mapping the LDA or GGA Hamiltonian onto the Wannier basis.  Hopping parameters obtained within GGA calculations in different codes are compared in Table~\ref{tab:Hoppings}. In the main text, we use the results of LDA calculations with spin-orbit coupling.

\par The on-site Coulomb, intersite Coulomb and intersite exchange interactions were calculated within constrained random phase approximation (cRPA) in the basis of Wannier functions~\cite{rpa1, rpa2}:
\begin{equation}
\begin{aligned}
U_{i}&=\int d\mathbf{r}d\mathbf{r}'w^{\dagger}_{i}(\mathbf{r})w^{\phantom{\dagger}}_{i}(\mathbf{r})U(\mathbf{r},\mathbf{r'})w^{\dagger}_{i}(\mathbf{r}')w^{\phantom{\dagger}}_{i}(\mathbf{r}') ,\\
V_{ij}&=\int d\mathbf{r}d\mathbf{r}'w^{\dagger}_{i}(\mathbf{r})w^{\phantom{\dagger}}_{i}(\mathbf{r})U(\mathbf{r},\mathbf{r'})w^{\dagger}_{j}(\mathbf{r}')w^{\phantom{\dagger}}_{j}(\mathbf{r}'), \\
J_{ij}^{F}&=\int d\mathbf{r}d\mathbf{r}'w^{\dagger}_{i}(\mathbf{r})w^{\phantom{\dagger}}_{j}(\mathbf{r})U(\mathbf{r},\mathbf{r'})w^{\dagger}_{j}(\mathbf{r}')w^{\phantom{\dagger}}_{i}(\mathbf{r}'),\\
\end{aligned}
\label{eq:coulomb}
\end{equation}
\noindent where $U(\mathbf{r},\mathbf{r}')$ is the partially screened Coulomb interaction defined for the target bands:
\begin{equation}
U=W\big(1+P_{d}W\big)^{-1} \qquad W=\big(1-Pv\big)^{-1}v,
\end{equation}
\noindent $v$ and $W$ are the bare and fully screened Coulomb interactions, respectively; $P$ is the total polarization given by all states, and $P_{d}$ is the partial polarization including transitions between the target bands only. The corresponding parameters were computed from LDA calculations in \texttt{VASP}~\cite{rpa3} and are presented in Table~\ref{tab:cRPA}. We used $20\times20\times1$ and $4\times16\times1$ $k$-point meshes for LaBr$_{2}$ and  La$_{2}$Br$_{5}$, respectively.  

\par Matrix elements of the bare Coulomb and exchange interactions in the basis of Wannier functions were also calculated by direct integration of Eq.~(\ref{eq:coulomb}) with $v(\mathbf{r},\mathbf{r}')=e^{2}/|\mathbf{r}-\mathbf{r}'|$ in real space and are in good agreement with \texttt{VASP} calculations.

\begin{table}
\centering
\caption{Hopping parameters (in meV) between neighbouring anionic electrons in LaBr$_{2}$ and La$_{2}$Br$_{5}$, as obtained from GGA calculations without spin-orbit coupling in \texttt{QE}, \texttt{ELK}, and \texttt{VASP}.}
\begin{ruledtabular}
\begin{tabular}{crrrcrrr}
\multicolumn{4}{c}{LaBr$_2$}  &  & \multicolumn{3}{c}{ La$_2$Br$_5$ }   \\ 
 $t_{ij}$ &   \texttt{QE} &  \texttt{ELK}  &  \texttt{VASP}  &  &  $t_{ij}$ &   \texttt{QE}  &  \texttt{VASP}  \vspace{0.1cm} \\
  \cline{1-4} \cline{6-8}  \vspace{-0.2cm} \\ 
  1$^{st}$    &      14.1  &    11.0     &  11.7    &     &    1$^{st}$     &   $-14.7$   & $-18.8$   \\
  2$^{nd}$  &      63.0  &    66.8   &   62.2  &     &    2$^{nd}$   &    $-2.0$    & $-2.4$     \\
  3$^{rd}$   &     $-14.2$  &   $-17.9$   & $-13.9$   &     &    3$^{rd}$   &     11.6    &  11.8    \\
  4$^{th}$   &      $-2.7$   &     $-2.1$   & $-2.7$     &     &    4$^{th}$    &   $-41.1$    & $-40.6$   \\
 \end{tabular}
\end{ruledtabular}
\label{tab:Hoppings}
\end {table}

\par $\qquad$
\par $\qquad$
\par $\qquad$
\par $\qquad$

\begin{table}[!h]
\centering
\caption{Results of cRPA calculations for LaBr$_{2}$ and La$_2$Br$_{5}$. Nearest neighbours with the distance $d$ are shown in figures 4b and 5b of the main text. Here, $J^{\mathrm{bare}}$ and $J^{\mathrm{full}}$ stand for the bare and fully screened exchange  interactions (as the corresponding matrix elements of $v$ and $W$ in the Wannier basis). $^{*}$This number is derived by taking the ratio $J^{F}_{}/J^{\mathrm{bare}}_{}$ from first nearest neighbours.
}
\begin{ruledtabular} 
\begin{tabular}{ccccccccccc}
\multicolumn{5}{c}{LaBr$_2$}  &  & \multicolumn{5}{c}{ La$_2$Br$_5$ }   \\ 
 \cline{1-5} \cline{6-11}  
 \vspace{-0.3cm} \\
\multicolumn{5}{c}{ On-site Coulomb (eV)} &  &  \multicolumn{5}{c}{On-site Coulomb (eV)} \\
\multicolumn{5}{c}{$v_{i}=6.071$ $\qquad$ $W_{i}=0.237$ $\qquad$ $U_{i}=1.536$} &  &  \multicolumn{5}{c}{$v_{i}=5.476$ $\qquad$ $W_{i}=0.769$ $\qquad$ $U_{i}=2.041$}\\ 
\cline{1-5} \cline{6-11}\\    
& $\,\,\,\,\,$ 1$^{st}$ & $\,\,\,\,\,$ 2$^{nd}$ & $\,\,\,\,\,$ 3$^{rd}$ & $\,\,\,\,\,$ 4$^{th}$  &  & & $\,\,\,\,\,$ 1$^{st}$ & $\,\,\,\,\,$ 2$^{nd}$ & $\,\,\,\,\,$ 3$^{rd}$ & $\,\,\,\,\,$ 4$^{th}$   \\
$d$ (\AA) & $\,\,\,\,\,$ 4.099 & $\,\,\,\,\,$ 7.099 & $\,\,\,\,\,$ 8.198 & $\,\,\,$ 10.844     &  &  $d$ (\AA) & $\,\,\,\,\,$ 4.301 & $\,\,\,\,\,$8.127 & $\,\,\,\,\,$ 8.379 & $\,\,\,\,\,$ 8.601 \\
\\
\multicolumn{5}{c}{Intersite Coulomb (eV)}  &  &  \multicolumn{5}{c}{Intersite Coulomb (eV)} \vspace{0.1cm} \\
$v_{ij}$ & $\,\,\,\,\,$ 3.352 & $\,\,\,\,\,$ 1.964 & $\,\,\,\,\,$ 1.652 & $\,\,\,\,\,$1.112     &  &  $v_{ij}$ & $\,\,\,\,\,$ 3.122 & $\,\,\,\,\,$ 1.664 & $\,\,\,\,\,$ 1.602 & $\,\,\,\,\,$ 1.574  \\
$W_{ij}$ & $\,\,\,\,\,$ 0.035 & $\,\,\,\,\,$ 0.028 & $\,\,\,\,\,$ 0.025 & $\,\,\,$  0.018   &  & $W_{ij}$ & $\,\,\,\,\,$ 0.496 & $\,\,\,\,\,$ 0.319 & $\,\,\,\,\,$ 0.386 & $\,\,\,\,\,$ 0.482 \\ 
$V_{ij}$ & $\,\,\,\,\,$ 0.717 & $\,\,\,\,\,$ 0.612 & $\,\,\,\,\,$ 0.431 & $\,\,\,$ 0.260       &  & $V_{ij}$ & $\,\,\,\,\,$ 1.181 & $\,\,\,\,\,$ 0.828 & $\,\,\,\,\,$ 0.648 & $\,\,\,\,\,$ 0.730 \\ 
\\
\multicolumn{5}{c}{Intersite exchange (meV)}  &  &  \multicolumn{5}{c}{Intersite exchange (meV)} \vspace{0.1cm} \\
$J^{\mathrm{bare}}_{ij}$ & $\,\,\,\,\,$ 44.2 & $\,\,\,\,\,$ 5.6 & $\,\,\,\,\,$ $1.4$ & $\,\,\,$  $0$  &  & $J^{\mathrm{bare}}_{ij}$ & $\,\,\,\,\,$ 36.4 & $\,\,\,\,\,$ 1.2 & $\,\,\,\,\,$ $0.7$ & $\,\,\,\,\,$ $9.9$ \\
$J^{\mathrm{full}}_{ij}$ & $\,\,\,\,\,$ 16.5 & $\,\,\,\,\,$ 1.6 & $\,\,\,\,\,$ $-$ & $\,\,\,$ $-$       &  & $J^{\mathrm{full}}_{ij}$ & $\,\,\,\,\,$ 13.2 & $\,\,\,\,\,$ $-$ & $\,\,\,\,\,$ $-$ & $\,\,\,\,\,$ $-$ \\ 
$J^{F}_{ij}$ & $\,\,\,\,\,$ 20.3 & $\,\,\,\,\,$ 4.4 & $\,\,\,\,\,$ $-$ & $\,\,\,$ $-$       &  & $J^{F}_{ij}$ & $\,\,\,\,\,$ 25.0 & $\,\,\,\,\,$ $-$ & $\,\,\,\,\,$ $-$ & $\,\,\,\,\,$ $\,\,\,6.7^{*}$ \\ 
\end{tabular}
\end{ruledtabular}
\label{tab:cRPA}
\end {table}
\vspace{2.0cm}
\par $\qquad$
